\documentstyle[11pt]{article}
\input{epsf}
\begin{document}

\title{Bounds on the existence of neutron rich nuclei\\  
 in neutron star interiors\footnote{Invited talk 
 presented by P. Haensel at the international 
conference ``Nuclear Physics Close to the Barrier'', 
Warsaw, Poland,  30.06 - 4.07.1998.}
}

\author{{\bf F.\ Douchin}$^{1}$, {\bf P.\ Haensel}$^{1,2}$\\ 
        $^1${\it Centre de Recherche Astronomique de Lyon, } \\
        {\it Ecole Normale Sup{\'e}rieure de Lyon, 
          46, all{\'e}e d'Italie,}\\
        {\it 69364 Lyon,  France}\\
        $^2${\it N.\ Copernicus Astronomical Center,} \\
        {\it Polish Academy of Sciences, Bartycka 18,}\\
        {\it 00-716 Warszawa, Poland}}
\date{${June~30th,~1998}$}
\maketitle

\def\la{\;\raise0.3ex\hbox{$<$\kern-0.75em\raise-1.1ex\hbox{$\sim$}}\;}
\def\ga{\;\raise0.3ex\hbox{$>$\kern-0.75em\raise-1.1ex\hbox{$\sim$}}\;}
\newcommand{\dd}{\mbox{d}}                     
\begin{abstract}
 We address the question concerning the maximum 
density, 
$\rho^{\cal N}_{\rm max}$, 
at which nuclei (and more generally -- nuclear structures) 
 can exist in neutron 
star interiors. 
An absolute upper bound to  
$\rho^{\cal N}_{\rm max}$ is obtained using the bulk approximation, in which 
surface  and Coulomb effects are neglected. 
A very good approximation to  
$\rho^{\cal N}_{\rm max}$ is given by the threshold for the instability 
of a uniform  $npe$ plasma  with respect to  density modulations; 
this threshold is calculated using the Extended Thomas-Fermi approximation 
for the Skyrme energy functionals. 
 For recent SLy Skyrme
forces, which are particularly suitable for the description of very neutron 
rich nucleon systems, one gets   
 $\rho^{\cal N}_{\rm max}=0.08~{\rm fm^{-3}}$; at this density protons 
constitute only 4\% of nucleons.   
\end{abstract}
\vskip 2mm
PACS numbers: 97.60. Jd, 21.65. +f, 95.30. Cq\par
\vskip 3mm
\vskip 5mm

\section{Introduction}
Neutron drip instability limits the neutron excess of neutron--rich nuclei 
which can be formed  in laboratory to $\delta=(N-Z)/A<\delta_{\rm n-drip}
\simeq 0.3$. However, this limitation is no longer valid for nuclei in the 
interiors of neutron stars, where the density varies from a few 
${\rm g~cm^{-3}}$ near the stellar surface  to more than $10^{15}~{\rm 
g~cm^{-3}}$ near the star center. Above $10^{4}~{\rm g~cm^{-3}}$, atomic  
structures are crushed, and electrons form an essentially  uniform Fermi 
gas. Standard scenario of neutron star formation in gravitational collapse 
of a massive stellar core (or of a mass accreting white dwarf) 
predicts nuclear composition corresponding to 
the state of complete thermodynamic equilibrium 
 (minimum of the free energy per nucleon).  
 Neutron excess parameter of nuclei, immersed in electron gas, increases 
with increasing density (i.e., increasing depth below stellar surface).   
Beta decay of neutron rich nuclei, which would be unstable in vacuum,  
 is  blocked via Pauli exclusion 
principle due to the presence of dense, degenerate  
electron gas. 
Up to the density $4-6\times 10^{11}~{\rm g~cm^{-3}}$, neutron star matter 
consists of nuclei immersed in dense electron gas. At higher densities, 
neutrons  start to populate continuum states, forming a degenerate 
neutron gas. The presence of an outer degenerate 
neutron gas influences the properties 
of nuclei by:  {\it a)} blocking (due to  Pauli exclusion principle) 
neutron emission from nuclei, {\it b)} exerting  
 pressure on nuclei (i.e. compressing them ), and  {\it c)} 
modifying (lowering) nuclear surface energy. 

Further increase of density is accompanied by an increase of the fraction 
of volume occupied by nuclei and a simultaneous decrease of proton fraction 
in neutron star matter (and in nuclei). 
In classical terms, nucleon component of neutron 
star matter 
consists there of two coexisting nucleon fluids: denser  one in the interior of 
nuclei, and the less dense outer neutron gas (at highest densities the 
less dense nucleon fluid  can contain some admixture of protons, see 
Section 2). At 
some density $\rho^{\cal N}_{\rm max}$ the  two-fluid phase becomes unstable
with respect to transition into a uniform, electrically neutral 
  $npe$ plasma, composed mostly 
of neutrons, with a few percent admixture of electrons and protons. 
The density  $\rho^{\cal N}_{\rm max}$ is the maximum density, at which 
nuclei can exist in the neutron star interiors; it turns out 
to be significantly 
lower than normal nuclear saturation density $\rho_0=0.16~{\rm fm^{-3}}$, 
typically $\sim {1\over 2}\rho_0$ (see Section 3).  
 Before  $\rho^{\cal N}_{\rm max}$ is reached, the interplay 
between Coulomb and surface effects can lead to the appearance of 
unusual  shapes of `nuclei' (rods, plates, tubes, bubbles, ...). However, 
both the theoretical 
value of  $\rho^{\cal N}_{\rm max}$  and the actual shape of 
 `nuclei' (the more appropriate term would be `nuclear structures') at 
highest densities,   
depend on the assumed effective nuclear hamiltonian, used in many--body 
calculations  \cite{LRP93,Oyam93,
PethRav95,Sumiy95,Cheng97}. For some effective N-N interactions, neutron 
star matter contained spherical nuclei down to $\rho^{\cal N}_{\rm max}$ 
(SkM force in \cite{LRP93}, all relativistic mean field models in 
\cite{Cheng97}). For  other models of effective  N-N forces, spherical 
nuclei were replaced  by  a sequence of 
cylindrical ones (rods), flat (plates), 
cylindrical holes in nuclear matter  filled by neutron gas (tubes), 
and finally  spherical holes in nuclear matter filled by neutron gas 
(bubbles), before disappearence of nuclear structures at 
 $\rho^{\cal N}_{\rm max}$ \cite{LRP93,Oyam93,
Sumiy95,PethRav95}.   

Let us stress that up to about $\rho_0$ the elementary constituents of 
neutron star matter are still the same as those of terrestrial matter: 
 $n$, $p$, $e$ (see, e.g., \cite{ShapTeuk83}). However, under extremal 
conditions prevailing in neutron star interior, the {\it structures} 
these constituents form can be dramatically different from those familiar
from terrestrial ones. 

The quantity $\rho^{\cal N}_{\rm max}$ has an important astrophysical 
meaning: it determines the bottom boundary of the neutron star {\it crust},   
 an outer envelope, in which nuclei form a periodic crystal lattice due 
to longe range Coulomb interactions. In the case of unusual nuclei, 
one deals rather with two dimensional (rods, tubes) or even one dimensional 
(plates) liquid crystals \cite{LRP93,Oyam93,PethRav95,PetPotek98}. The solid 
crust plays an important role in the evolution and dynamics of neutron stars 
(see, e.g. \cite{ShapTeuk83}).  

In view of the fact, that the bottom layers of neutron star crust involve 
nucleon matter with an extremely large neutron excess ($\delta\simeq 0.9$), 
 an appropriate choice of effective nuclear 
hamiltonian is of crucial importance.  
Recently, a new  set of the Skyrme-type effective N-N interactions 
SLy (Skyrme Lyon) has 
been derived, based on  an approach which is particularly appropriate, as 
far as the applications to a  neutron rich matter are concerned 
\cite{Chabanat1,Chabanat2}. Relevant additional experimental items
concerning nuclei with large neutron excess, isovector effective  
masses, constraints of spin stability, and requirement of consistency 
with the realistic UV14+VIII equation of state (EOS)
  of dense neutron matter of Wiringa et al. \cite{WFF88}
  for $\rho_0\le \rho\le 1.5~{\rm fm^{-3}}$, were included   into a   
 fitting procedure for the SLy forces parameters. 
 In the present paper we calculate 
$\rho^{\cal N}_{\rm max}$ 
 using  the SLy models of effective 
N-N interaction, and compare our results with those obtained using 
older Skyrme-type 
forces, Sk1$^\prime$ and SkM$^*$, used frequently in astrophysical 
applications. The parameters of the SLy forces, 
used in the present calculations, 
together with those of the  SkM$^*$ and Sk1$^\prime$ models,  
are given in Table 1. 
The SLy4 is a basic SLy force; the  SLy7 model has been obtained 
 following the most ambitious fitting procedure, in which 
spin-gradient  
terms and center of mass correction term  were both included in the Skyrme 
energy functional \cite{Chabanat2}.   

At the densities of interest, matter in the interior of a neutron 
star which is more than one year old, is strongly degenerate, and thermal 
contributions to thermodynamic quantities can be safely neglected. 
In what follows, we will consider the properties of neutron star matter 
in the $T=0$ approximation; dense matter will be assumed to be 
in its ground state (it is then called `cold catalyzed matter').
\vskip 3mm
%
\begin{tabular}{ccccc}
\multicolumn{5}{c}{Table 1}\\
\multicolumn{5}{c}{Parameter values of the Skyrme forces}\\
&&&&\\
\hline\hline
&&&&\\
force & SLy4 &  SLy7  &  SkM$^*$ & Sk1$^\prime$\\ 
&&&&\\
\hline
&&&&\\
 $t_0$~ $({\rm MeV~fm^3})$ & -2488.91   & -2482.41    & -2645.0 & -1057.3    \\ 
$t_1$~ $({\rm MeV~fm^5})$ & 486.82    &  457.97   & 410.0   & 235.9     \\
$t_2$~$({\rm MeV~fm^5})$ & -546.39      &  -419.85    & -135.0 &   -100.00   \\
$t_3$~$({\rm MeV~{fm}^{3+3\sigma}})$ & 13777.0& 13677.0  & 15595.0  & 14463.5 \\
 $\sigma$  &  $1\over 6$ & $1\over 6$  &  $1\over 6$  &  1 \\
 $x_0$  & 0.834   & 0.846  & 0.09  &  0.2885 \\
 $x_1$  & -0.344  & -0.511   & 0  & 0  \\
 $x_2$  &  -1.000 & -1.000  & -1.000  & 0  \\
 $x_3$  & 1.354   & 1.391  &  0  &  0.2257 \\
$W_0$~$({\rm MeV~fm^5})$ &  123.0  & 126.0   & 130.0   & 120.0   \\
&&&&\\
\hline
\end{tabular}
\vskip 3mm
 
In Section 2, we present the calculation of 
 $\rho^{\cal N}_{\rm max}$ in the bulk 
approximation. Threshold for the instability of homogeneous $npe$ matter 
with respect to density modulations (i.e., formation of `nuclear structures') 
is calculated in Section 3. Properties of very neutron rich nuclei in 
the bottom layers of neutron star crust, and  dependence on 
the effective N-N interaction model,  are discussed in Section 4. 
Concluding remarks are presented in 
Section 5.
   
\section{Bound for the existence of nuclei:  the bulk approximation}
Above neutron drip density, $\rho_{\rm n-drip}$, neutron star crust 
is  a two-phase nucleon system, the denser nucleon 
phase (fluid)  `i' residing inside nuclei and 
the less dense `o' one forming a gas 
outside nuclei. Both nucleon fluids are permeated by an essentially homogeneous 
electron gas, which ensures overall charge neutrality.   
The shape of nuclei, i.e. that  of the `i-o' interface,  results  from 
the balance of the surface and Coulomb terms in the total energy of the system. 

The bulk approximation consists in neglecting the Coulomb and surface 
effects. At given mean nucleon density $\rho$, nucleons are present 
in general in both  of the coexisting `i' and `o' fluids, 
characterized by the constant densities 
$\rho_{n{\rm i}}$, 
$\rho_{p{\rm i}}$, 
$\rho_{n{\rm o}}$, and above proton drip density, $\rho_{\rm p-drip}$, also  
$\rho_{p{\rm o}}$. 
The equilibrium between 
the `i' and `o' fluids, ensured by the (strong) N-N interaction, implies 
the equality  of the chemical potentials of nucleons, and the equality 
of nucleon pressures,
\begin{eqnarray}
\mu_{n{\rm i}}=\mu_{n{\rm o}},~~~~~~
\mu_{p{\rm i}}=\mu_{p{\rm o}}~,
~~~~~P_{N{\rm i}}=P_{N{\rm o}}~,
\label{eq.bulk}
\end{eqnarray}
where the label $N$ indicates the nucleon contribution to the matter 
pressure, and the condition on the proton chemical potentials applies 
only above the proton-drip density, $\rho_{\rm p-drip}$. At a given $\rho$,  
 Eqs. (\ref{eq.bulk}) enables determination of 
$\rho_{n{\rm i}}$, 
$\rho_{p{\rm i}}$, 
$\rho_{n{\rm o}}$, and for $\rho>\rho_{\rm p-drip}$, also  
$\rho_{p{\rm o}}$. 

Let us denote the volume fraction occupied by the `i' fluid by $u$. 
At a given mean nucleon density $\rho=u\rho_{\rm i} + (1-u)\rho_{\rm o}$, 
the total energy density is given by
\begin{equation}
E=uE_{N{\rm i}} + (1-u)E_{N{\rm o}} + E_e~,
\label{E.bulk}
\end{equation}
where $E_e$ is the electron energy density, $E_e(\rho_e)$. Beta equilibrium 
within the   $npe$ matter implies relation between the chemical potentials 
of neutrons, protons and electrons,  
\begin{equation}
\mu_n=\mu_p + \mu_e~, 
\label{beta.eq}
\end{equation}
while requirement of the overall charge neutrality leads to 
\begin{equation}
\rho_e=u\rho_{p{\rm i}}+ (1-u)\rho_{p{\rm o}}~.
\label{ch.neutr}
\end{equation}
Eqs. (\ref{eq.bulk},\ref{beta.eq},\ref{ch.neutr})  
determine completely the equilibrium of a  two-fluid $npe$ matter 
at a given nucleon density $\rho$. 
\begin{figure}
\begin{center}
\epsfysize=16cm  \epsfbox{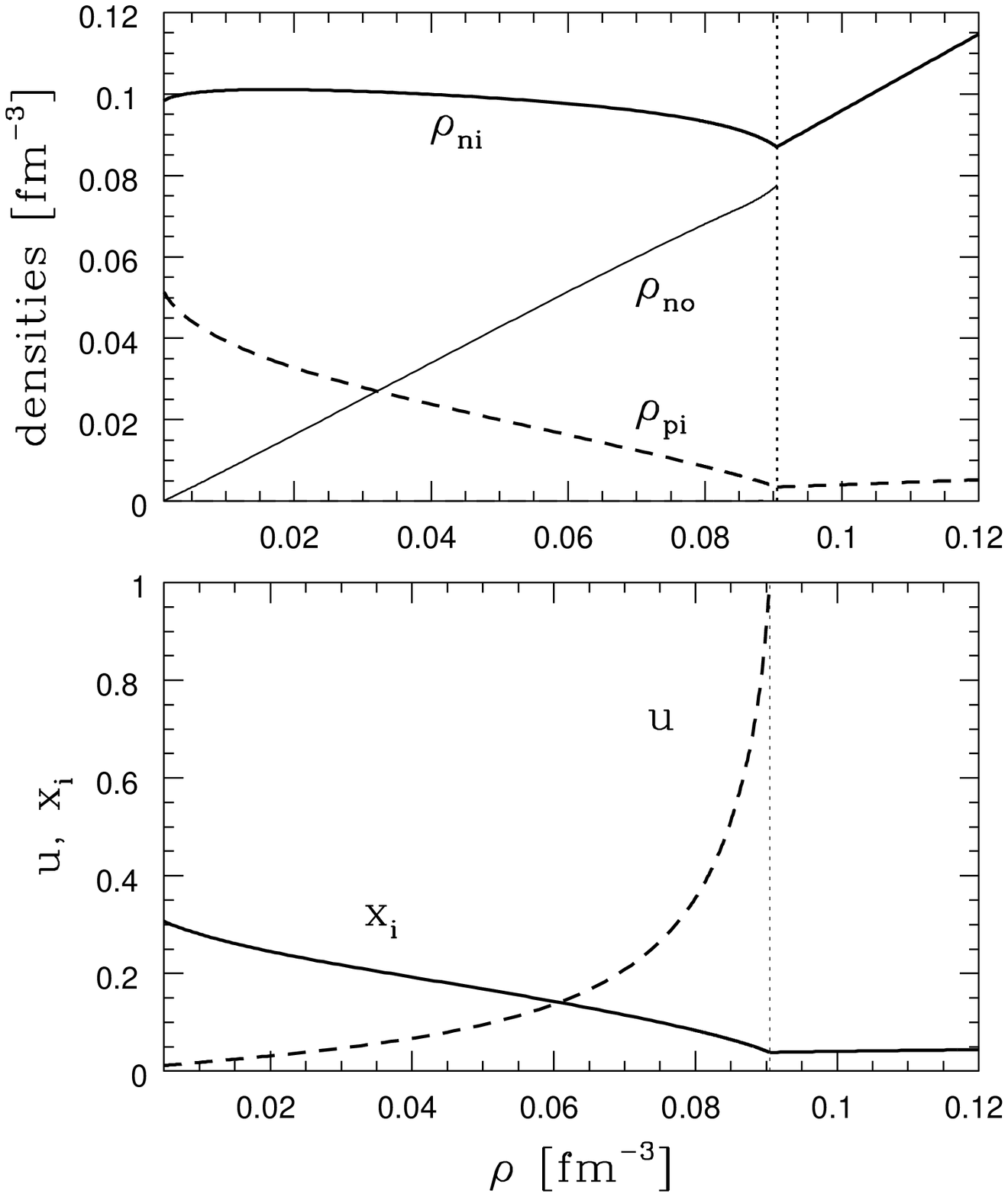}
\end{center}
{
Figure 1. Composition and the two-fluid - one-fluid phase transition 
 in the bulk approximation, for the 
SLy4 force. Two-fluid phase to the left, 
one-fluid phase to the right of the dotted line representing the 
bottom boundary of the crust in the bulk approximation.
}
\end{figure}

The two-fluid phase  
 is stable at a given $\rho$ if 
its energy per nucleon is lower than that in the uniform  
(one-fluid) phase (i.e., that of a uniform  $npe$ matter). 
This stability condition breaks down at a density  
$\rho_{2\leftrightarrow 1}$. 
 Calculations show that approaching  
$\rho_{2\leftrightarrow 1}$ 
from the lower density (two-fluid) side corresponds to 
 $u\longrightarrow 1$, i.e. to the `i' (denser) fluid filling 
the whole volume  \cite{PethRL95}. 
 Therefore the $2\leftrightarrow 1$ 
transition is a continuous one, with no density jump (our results 
for the SLy4 force are shown in Fig. 1).
Actually, Coulomb and surface effects would 
increase the energy per nucleon in the 
two-fluid phase, as compared to the plain bulk approximation. 
Therefore, the real transition into a uniform  $npe$ matter 
occurs at density lower than  
$\rho_{2\leftrightarrow 1}$, 
 so that $\rho_{2\leftrightarrow 1}$ 
 is thus a robust {\it upper bound} 
to $\rho^{\cal N}_{\rm max}$. 

\begin{tabular}{ccccc}
\multicolumn{5}{c}{Table 2}\\
\multicolumn{5}{c}{Neutron drip and proton drip densities,  
and proton fraction}\\ 
\multicolumn{5}{c}{ and nucleon density at the  
 2-fluid$\leftrightarrow$1-fluid transition,}\\
\multicolumn{5}{c}{ calculated 
in the bulk approximation}\\
\multicolumn{5}{c}{}\\
\hline\hline
&&&&  \\
force &  $\rho_{n-{\rm drip}}$ & $\rho_{p-{\rm drip}}$ & 
$\rho_{2\leftrightarrow 1}$ &
  $x_{2\leftrightarrow 1}$ \\
&&&&   \\
  &  $({\rm fm^{-3}})$ &
    $({\rm fm^{-3}})$ &
  $({\rm fm^{-3}})$ & (\%)  \\
&&&&  \\
\hline
&&&& \\
${\rm SkM^*}$  &$7.55\times 10^{-4}$ & 0.0781 & 0.0866 & 3.13  \\
&&&& \\
${\rm Sk1'}$ & $8.09\times 10^{-4}$ & 0.1066  & 0.1088 & 3.71  \\
&&&&\\
${\rm SLy4}$ & $7.34\times 10^{-4}$ & 0.0853 & 0.0905 & 3.84  \\
&&&&\\
${\rm SLy7}$ & $7.21\times 10^{-4}$ & 0.0836 & 0.0889 & 3.80   \\
&&&&\\
\hline
\end{tabular}
\vskip 3mm
Numerical values of $\rho_{2\leftrightarrow 1}$ for four  Skyrme 
forces are shown in Table 2, where we show also  the proton-drip 
densities, and the proton fractions at the  $2\leftrightarrow 1$ 
transition point, $x_{2\leftrightarrow 1}$ 
(bulk instability of two-fluid phase with Skyrme forces 
SkM and Sk1$^\prime$ was studied in \cite{PethRL95}). . 
 As far as the value of $\rho_{2\leftrightarrow 1}$ 
is concerned, both SLy models yield values close to $0.09~{\rm fm^{-3}}$, 
 intermediate 
between those corresponding to the Sk1$^\prime$
and SkM$^*$ forces, which yield  $0.11~{\rm fm^{-3}}$ and 
 $0.087~{\rm fm^{-3}}$, respectively.  
 Proton drip takes place in a narrow 
interval of densities close to  
$\rho_{2\leftrightarrow 1}$. At the  $2\leftrightarrow 1$ transition point, 
protons constitute less than 4\% of nucleons.

\vskip 3mm
\section{Instability of a uniform  $npe$ matter and 
existence of nuclear structures}
An instability of a spatially uniform  state of the $npe$ matter with 
respect to density modulations signals the appearence of nuclear structures 
in the true ground state of the $npe$ system. Let us consider first 
a $npe$ matter at 
nucleon density $\rho>\rho_0$, which is sufficiently high to exlude any 
possibility of existence of nuclear structures. By decreasing $\rho$, 
we will eventually find the threshold density, below which uniform $npe$ 
matter is unstable with respect to density modulations. This threshold density
 turns out a lower bound (and a rather good approximation) 
to  $\rho^{\cal N}_{\rm max}$ 
\cite{PethRL95,PethRav95}. 

The ground state of uniform  $npe$ matter of a  
 given nucleon density  $\rho$ corresponds to the minimum 
of the energy density $E(\rho_n,\rho_p,\rho_e)=E_0$,  
 calculated  
at  a fixed nucleon density 
 $\rho=\rho_p+\rho_n=\rho$, under the condition of charge neutrality  
 $\rho_e=\rho_p$ (all  densities being  {\it assumed}  constant 
in space).   
Minimisation  implies the beta equilibrium relation 
between the chemical potentials of matter constituents, 
$\mu_n=\mu_p + \mu_e$. 
 This relation ensures vanishing of the first variation of  
$E$ implied by 
 small perturbations $\delta\rho_{j}({\bf r})$ (where 
 $j=n,p,e$). 
  However, this 
 does not guarantee the {\it stability} of the spatially 
homogeneous state of the 
$npe$ matter, which requires that the second  variation  of  
$E$ (quadratic in $\delta\rho_j$)   
 be positive. 

The energy per unit volume, $E$, of a slightly nonuniform $npe$ matter, 
can be decomposed into nucleon, electron, and 
Coulomb components,
\begin{equation}
E=E_N + E_e  + E_{\rm Coul}~.
\label{decompE}
\end{equation}
\parindent 0pt
The nucleon contribution to $E$ can be expressed in terms of the energy 
density\par
 ${\cal E}_N(\rho_n,\rho_p,\nabla\rho_n,\nabla\rho_p)$, 
\parindent 21pt
obtained 
from the Skyrme  models of the effective nucleon hamiltonian,   
 so that for a perturbed, slightly  
spatially inhomogeneous state 
we get the energy functional 
\begin{equation}
E_N={1\over V}\int \dd {\bf r}
{\cal E}_N[\rho_n({\bf r}),\rho_p({\bf r}),
\nabla\rho_n({\bf r}),\nabla\rho_p({\bf r})]~. 
\label{E_n}
\end{equation}
The expression for ${\cal E}_N$ has been calculated using the 
semi-classical Extended 
Thomas-Fermi (ETF) treatment of the kinetic  and 
 the spin-orbit energy densities \cite{BrackJC76}.  
Assuming that the spatial 
gradients are small, one  keeps only the quadratic gradient terms in the ETF 
expressions. This approximation will be justified by 
the fact that characteristic 
wavelengths of  perturbations turn out to 
 be much larger then the 
internucleon distance (see below). 
 With such approximations, 
 the change of the energy (per unit volume) 
 implied 
by the  density perturbations can be expressed, keeping only second order 
terms, as  \cite{BBP,PethRL95},  
\begin{equation}
E-E_0= {1\over 2}\int{{\rm d}{\bf q}\over (2\pi)^3} 
 F_{ij}({\bf q})\delta\rho_i({\bf q})\delta\rho_j({\bf q})^*~,
\label{E-E_0}
\end{equation}
where  Fourier representation
\begin{equation}
\delta\rho_j({\bf r})=\int{{\rm d}{\bf q}\over (2\pi)^3} 
\delta\rho_j({\bf q}){\rm e}^{{\rm i}{\bf q}{\bf r}}~.  
\label{rho.Fourier}
\end{equation}
has been used. The hermitian $F_{ij}({\bf q})$ matrix determines the 
stability of the uniform state of equilibrium of the $npe$ matter with 
respect to the spatially periodic perturbations of the densities 
of wavevector ${\bf q}$. 
Due to the isotropy of the homogeneous equilibrium state of 
the $npe$ matter, $F_{ij}$ depends only on $\vert {\bf q}\vert=q$.  
In the case of Skyrme forces,  
 the matrix elements 
$F_{ij}$ can be calculated analytically, as explicit functions 
of the equilibrium densities and $q$, and are composed of compression 
(local), curvature (gradient), and Coulomb components \cite{BBP,PethRL95}. 

The condition for the $F_{ij}$ matrix to be positive definite turns out 
to be equivalent to the requirement that the determinant of the $F_{ij}$ 
matrix be positive \cite{LRP93}. At each density $\rho$, 
one has thus to check the 
condition 
${\rm det}[F_{ij}(q)]>0$.
 Let us start with some $\rho>\rho_0$, at which 
${\rm det}[F_{ij}(q)]>0$ for any $q$. By decreasing $\rho$, we find 
eventually a wavenumber $Q$ at which stability condition is violated 
for the first 
time; this happens at some density $\rho(Q)$. For $\rho<\rho(Q)$ homogeneous 
state is no longer the true 
ground state of the $npe$ matter since it is  unstable 
with respect to small periodic density modulations.

In contrast to the bulk approximation, in which constant densities 
were assumed {\it by construction}, the general energy functional 
used in this section allows for a consistent treatment of spatially 
inhomogeneous states of the $npe$ matter. The instability at $\rho(Q)$ 
signals  a  phase  transition with a loss of translational symmetry 
of the $npe$ matter. In principle, this could be a second order phase 
transition. However, the combination of  Coulomb and surface effects turns 
 out to be sufficiently strong to destabilize the $npe$ state  with an 
 infinitesimal density modulation, leading to a first-order phase transition 
into a state with finite amplitude density modulations. 
The real equilibrium 
phase transition (at constant pressure) 
 will thus  take place at a $npe$ matter density 
 $\rho_1$, slightly higher than  
 $\rho(Q)$. Homogeneous $npe$ matter of  density $\rho_1$
coexists with a spatially inhomogeneous phase exhibiting some nuclear 
structures,  of  a density $\rho_2$ slightly
  lower    than $\rho_1$. 
Therefore, $\rho(Q)$ is actually a {\it lower bound} on 
 $\rho_1$, but in view of the closeness of $\rho_1$ and $\rho_2$, 
it can be used as a rather good approximation of 
$\rho_{\rm max}^{\cal N}=\rho_2$.

\begin{tabular}{|c|ccc|ccc|c|}
\multicolumn{8}{c}{}    \\
\multicolumn{8}{c}{Table 3}    \\
\multicolumn{8}{c}{Threshold for the instability of the homogeneous $npe$ 
matter}\\
\multicolumn{8}{c}{with respect to  density modulations}\\
\multicolumn{8}{c}{}    \\
\hline\hline
&&&&&&&  \\
force &  & $\rho(Q)$ &  &  &  $Q$   &  & $ x(Q)$\\
&&&&&&&   \\
      &  & $({\rm fm^{-3}})$ & & & $({\rm fm^{-1}}$) & 
 & (\%) \\
&&&&&&& \\
\hline
&&&&&&& \\
      & a & b & c & a & b & c &  c \\
&&&&&&& \\
\hline
&&&&&&& \\
${\rm SkM^*}$ & 0.0738  & 0.0744 & 0.0754 & 0.277 & 0.284 & 0.299 &  2.79\\
&&&&&&& \\
${\rm Sk1'}$ & 0.0992 & 0.0993 & 0.1005 & 0.314 & 0.319 & 0.367 &  3.52 \\
&&&&&&&\\
${\rm SLy4}$ & 0.0781 & 0.0787 & 0.0794 & 0.262 & 0.271 & 0.281 & 3.57 \\
&&&&&&&\\
${\rm SLy7}$ & 0.0773 & 0.780 & 0.0786 & 0.270 & 0.280 & 0.292 & 3.55 \\
&&&&&&&\\
\hline
\end{tabular}
\vskip 0.3cm
\indent a: TF approximation without spin-orbit term. \hfill\\
\indent b: ETF without spin-orbit term. \hfill\\
\indent c: ETF including spin-orbit term.  \hfill\\
\vskip 3mm
Our numerical results for four Skyrme forces  are shown 
in Table 3 (\cite{DouchinH98}, the case of Sk1$^\prime$ force was 
studied previously, using slightly different approximations, 
 in \cite{PethRL95}).  
 Column ``a'' was obtained using  
standard TF approximation to  
nucleon kinetic energy densities, and neglecting spin-orbit  
contribution to ${\cal E}_N$.
Adding quadratic gradient 
terms in the ETF expansion of the nucleon energy densities \cite{BrackJC76}  
 (columns ``b'') increased 
only slightly the values of $\rho(Q)$ (by less than 1\%) as 
compared to the TF values. Notice that both ``a'' and ``b'' were 
obtained neglecting the spin-orbit term in ${\cal E}_N$. 
Finally, adding the spin-orbit term, calculated using the ETF 
approximation \cite{BrackJC76} (column ``c''), further increased $\rho(Q)$ by 
about 1\% with respect to the ``b'' ones  (from  0.8\% for SLy7 
to 1.5\% for Sk1$^\prime$). In the case of the critical wavenumber $Q$,  
corresponding  effects  are significantly higher.

For the ETF approximation to be correct, the value of the characteristic 
wavelength of critical density perturbations, $2\pi/Q$, must be significantly 
larger than the mean internucleon distance. Despite a small proton fraction, 
$2\pi/Q\sim 17-22~$fm is typically  four times higher than the distance 
between protons  $r_{pp}=(4\pi\rho/3)^{-1/3}$; 
for neutrons this ratio is typically about eight. 
%
\section{Nuclei in the bottom layers of the crust}
The actual structure of the bottom layer of neutron star crust should be 
calculated including Coulomb and surface effects. One usually considers 
a limited set of 
five possible nuclear shapes: spheres of i-phase embedded in o-gas, 
rods of i-phase in o-gas, equidistant plates of i-phase with 
o-gas between them, tubes in the i-matter filled with o-gas,  
and spherical bubbles of the o-gas in i-matter. For a given nuclear 
shape, the ground state of the matter is calculated using the Wigner-Seitz
 approximation, in which the real system is replaced by a set of 
non-interacting, electrically neutral cells,  each cell containing one 
nucleus. 
Wigner-Seitz cells are spherical, cylindrical or plate-like, 
depending on the symmetry of nuclear structure. The total volume of cells 
is equal to the volume of the real system.

\begin{figure}
\begin{center}
\epsfysize=12cm  \epsfbox{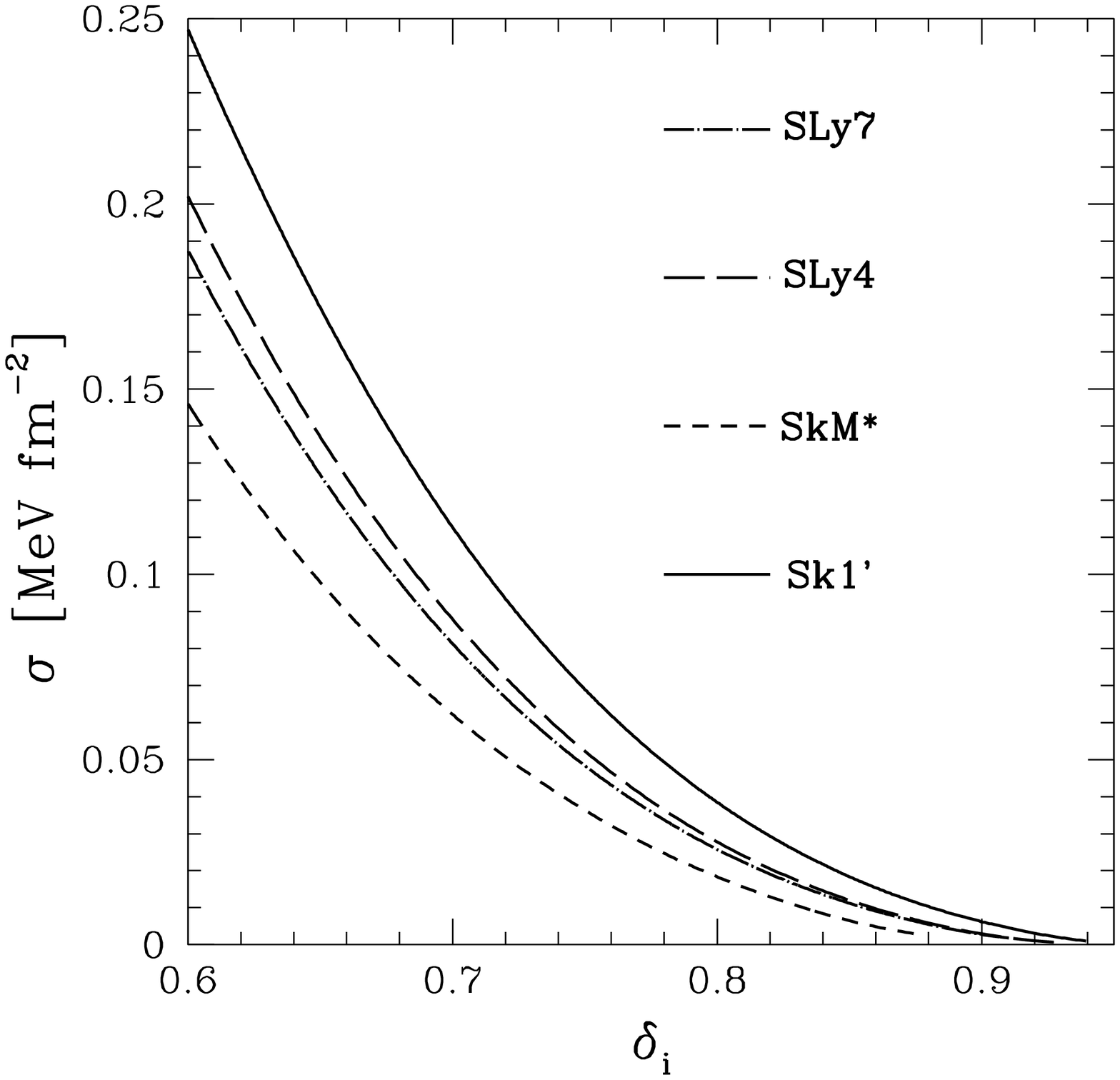}
\end{center}
{
Figure 2. Surface tension of the i-o phase interface versus 
neutron excess parameter $\delta_{\rm i}=1-2x_{\rm i}$  in the denser 
(i) phase far from the interface. Typical values of $\delta_{\rm i}$ 
($x_{\rm i}$) 
range from 0.6 (0.2) at 
$\rho=0.05~{\rm fm^{-3}}$ 
to 0.8 (0.1) 
at  
$\rho=0.08~{\rm fm^{-3}}$.  
}
\end{figure}

By comparing the enthalpies per nucleon 
$(E+P)/\rho$ at given $P$, one finds the actual shape of nuclei at this 
pressure $P$. 
Depending on the assumed effective N-N interaction, unusual nuclear shapes  
(starting with rods) may appear as early as at $\rho\simeq 5-6\times 
10^{-2}~{\rm fm^{-3}}$ 
(the case of FPS force of \cite{LRP93}, \cite{Oyam93}, 
\cite{Sumiy95}),
 or do not appear 
at all, spherical nuclei being present down to $\rho_2$ (SkM force case of  
\cite{LRP93}, \cite{Cheng97}). Elementary considerations
within the liquid drop model indicate, that at lower densities spherical 
shape is the only possible one. A possibility (or rather a necessity) 
of appearance of unusual nuclear shapes at highest densities 
results from the fact, that  
the fraction occupied by i-phase (i.e., nuclear matter), $u$, increases 
with increasing $\rho$, and above some limiting density  
the sum of the surface and Coulomb energies can be reduced  by changing 
nuclear shape from spherical to an unusual  one. 
 The main point 
is whether unusual  shape will appear {\it before} the transition into 
a homogeneous $npe$ matter. 

In general, the structure of the bottom layers 
of the neutron star crust may be expected to be rather sensitive to 
the behavior of the nuclear surface tension, $\sigma$,  in the relevant 
density interval ${1\over 3}\rho_0<\rho<{2\over 3}\rho_0$. Our results 
obtained within the ETF approximation for the Skyrme energy functional, 
shown in Fig. 2  \cite{DHsurf98}, 
visualize strong dependence of $\sigma$ at large 
neutron excess on the Skyrme force used [notice that all these forces 
lead to very similar values of $\sigma(\delta_{\rm i}=0)$, consistent 
with experimental value extracted from the liquid droplet model]. At 
$\delta_{\rm i}=0.8$ (which corresponds to the proton fraction $x_{\rm i} 
=0.1$), characteristic of $\rho\simeq 0.08~{\rm fm^{-3}}$, nuclear 
surface tension for SkM$^*$ model is only half of that obtained for the 
Sk1$^\prime$ one, which in turn is  30\% higher than our values  for
the SLy forces. We should remind, however, that  even at $\rho\simeq 
{1\over 3}\rho_0$ we have to deal with nuclei which are very far 
from the most neutron rich nuclei available  in terrestrial nuclear 
 physics, and therefore we should rely on 
the extrapolation of nuclear models to very high neutron excess. 
This may be visualized by our preliminary results obtained within 
the compressible liquid drop model 
for SLy4 force, shown in Table 4.
\vskip 2mm 
\begin{tabular}{cccccccc}
\multicolumn{8}{c}{Table 4}    \\
&&&&&&&\\
\multicolumn{8}{c}
{Examples of spherical nuclei in the 
bottom layers of neutron star crust}\\ 
\multicolumn{8}{c}
{with SLy4 force. 
Curvature corrections to surface energy are neglected.}\\ 
\multicolumn{8}{c}
{$A_{\rm cell}$ is the number of nucleons in 
Wigner-Seitz cell,  $u$ is the fraction}\\
\multicolumn{8}{c}
{of volume occupied by protons, 
$Z$ and $A$ are numbers of protons  and  nucleons}\\
\multicolumn{8}{c}
{in nuclei, $R_n$, $R_p$ are corresponding (equivalent) 
radii.}\\
&&&&&&&\\
\hline\hline
&&&&&&&  \\
 $\rho$ &   $\rho_{\rm o}/ \rho_{\rm i}$ & $u$  & $R_p$ & $R_n-R_p$  
     & $Z$ & $A$ & $A_{\rm cell}$ \\ 
&&&&&&& \\
 $({\rm fm^{-3}})$ & & & $({\rm fm})$ & $({\rm fm})$ &  & &  \\
&&&&&&& \\
\hline
&&&&&&& \\
 0.0160 & 0.0932 & 0.0154 & 5.56 & 0.83 & 28.2 & 133.2 &  751.3 \\
&&&&&&&\\
 0.0425 & 0.2819 & 0.0551 & 6.20 & 0.86 & 25.1 & 154.9 & 769.3 \\
&&&&&&& \\
0.0501 & 0.3441 & 0.0748  & 6.42 & 0.85 & 24.3 & 162.5   & 744.8 \\
&&&&&&& \\
0.0628 & 0.4584 & 0.1269  & 6.98  & 0.80 & 23.6 & 186.9  & 705.6 \\
&&&&&&& \\
&&&&&&&\\
\hline
\end{tabular}

%
\section{Conclusion}
Neutron star interiors are expected to contain extremely 
neutron rich nuclei (or more 
generally ``nuclear structures'', including those with unusual  
shapes), with neutron excess far beyond the laboratory neutron-drip limit. 
 However, no stable nuclear  structure can exist above a 
specific limiting density, which turns out to be significantly smaller than 
the saturation density of symmetric nuclear matter. For Skyrme forces SLy, 
which   are particularly suitable for applications 
in neutron star matter calculations, this ultimate density is  
about $0.08~{\rm fm^{-3}}$, the proton fraction at the limiting density 
being about 4\%.   

The procedure of construction of SLy forces make them suitable also for 
the calculation of equation of state of neutron star matter at supranuclear 
densities, up to $8\rho_0$ \cite{Chabanat1}. Therefore these effective N-N 
interactions can be used for a unified description of the whole neutron star, 
including its liquid interior. A low value of $\rho^{\cal N}_{\rm max}$ 
implies a small mass of neutron star crust. Consider a neutron star model of 
a ``canonical mass''  $1.4~M_\odot$ (measured masses of binary radio pulsars 
are quite close to this canonical value). For SLy7 force, the mass of the 
crust constitutes then only 1.3\% of stellar mass, while contribution of the 
crust to the total moment of inertia of neutron star is about 2.8\%
\cite{DouchinH98}. 

In spite of its small mass, the outer mantle 
of neutron star,  containing neutron rich nuclei - neutron star crust - 
 is of  paramount importance 
for star  evolution and dynamics. It insulates thermally neutron star 
surface from its liquid interior, and therefore plays an essential role 
in neutron star cooling. It can accumulate mechanical stresses, leading 
to instabilities crucial for the glitches in pulsar timing. Finally, 
neutron star crust can support, in contrast to the liquid interior, nonaxial 
deformations (``mountains'')  which, 
combined with rapid rotation, could make neutron star a 
source of continuous gravitational radiation. This gives a 
strong astrophysical 
motivation for theoretical studies of neutron rich nuclei in neutron star 
interiors. 
\begin{center}
                 {\bf ACKNOWLEDGMENTS}
\end{center}
We are grateful to G. Chabrier and J. Meyer
for helpful remarks. 
P.H. acknowledges the hospitality 
of ENS de Lyon during his visit in the theoretical astrophysics group 
of CRAL. 
This work was supported in part by the KBN grant No. 2 P03D 
014 13 and by the French CNRS/MAE program Jumelage Astronomie Pologne.  


\end{document}